\documentclass[pre,twocolumn,twoside,byrevtex,superscriptaddress,floatfix]{revtex4-1}

\usepackage{currfile}
\usepackage{fontawesome}

\usepackage{color}

\definecolor{olivegreen}{rgb}{0.33333,.41961,0.18431}
\definecolor{forestgreen}{rgb}{0.13333,.5451,0.13333}
\definecolor{lightgrey}{rgb}{0.7,0.7,0.7}
\definecolor{verylightgrey}{rgb}{0.90,0.90,0.90}
\definecolor{grey}{rgb}{0.5,0.5,0.5}

\usepackage{graphicx,epsfig,verbatim,enumerate}
\usepackage{amssymb,amsmath}
\usepackage{ifthen}

\newboolean{twocolswitch}
\newboolean{footnotes}
\newboolean{numbercites}

\usepackage{longtable}

\usepackage{mathtools}

\newcommand{\tavg}[1]{\langle#1\rangle}

\newcommand{\sindex}[1]{}
\newcommand{\nindex}[1]{}

\newcommand{\www}[1]{\url{#1}}

\newcommand{\Req}[1]{Eq.~(\ref{#1})}

\usepackage{lettrine}

\newcommand{\dee}[1]{\textnormal{d}#1}

\newcommand{\erdosrenyi}{Erd\"{o}s-R\'{e}nyi}

\newcommand{\edgeinfprob}{\rho}

\newcommand{\edgeinffreqbare}{f^{\textnormal{inf}}}
\newcommand{\edgeinffreq}[1]{f_{(#1)}^{\textnormal{inf}}}
\newcommand{\edgeinffreqplain}{\edgeinffreq{\cdot}}

\newcommand{\crit}{\textnormal{crit}}
\newcommand{\trig}{\textnormal{trig}}

\newcommand{\veck}{\vec{k}}

\newcommand{\infprob}{B}
\newcommand{\gainratio}{\mathbf{R}}

\newcommand{\bidmark}{\textnormal{u}}
\newcommand{\inmark}{\textnormal{i}}
\newcommand{\outmark}{\textnormal{o}}

\newcommand{\kplain}{k}

\newcommand{\kin}{k_{\inmark}}
\newcommand{\kout}{k_{\outmark}}
\newcommand{\kbid}{k_{\bidmark}}

\newcommand{\Probin}{P^{(\textnormal{\inmark)}}}
\newcommand{\Probout}{P^{(\textnormal{\outmark)}}}
\newcommand{\Probbid}{P^{(\textnormal{\bidmark)}}}

\newcommand{\seedset}{\mathcal{N}_{0}}

\newcommand{\alert}[1]{#1}
\newcommand{\alertb}[1]{#1}

\usepackage{cancel}

\newcommand{\rbone}{\textnormal{\faFilm}}
\newcommand{\rbtwo}{\textnormal{\faLightbulbO}}

\newcommand{\rboneng}{N_{\rbone}}
\newcommand{\rbtwong}{N_{\rbtwo}}

\setboolean{twocolswitch}{true}
\setboolean{footnotes}{false}
\setboolean{numbercites}{true}

\begin{document}

\title{\protect     A simple person's approach to understanding the contagion condition for spreading processes on generalized random networks\mbox{}\\
  \bigskip
  {\small\textnormal{
      To appear in ``Spreading Dynamics in Social Systems'';\\
      \vspace{-2pt}
      Eds. Sune Lehmann and Yong-Yeol Ahn, Springer Nature.
    }
  }
}

\author{
  \firstname{Peter Sheridan}
  \surname{Dodds}
}

\email{peter.dodds@uvm.edu}

\affiliation{
  Vermont Complex Systems Center,
  Computational Story Lab,
  the Vermont Advanced Computing Core,
  Department of Mathematics \& Statistics,
  The University of Vermont,
  Burlington, VT 05401.
  }

\date{\today}

\begin{abstract}
  \protect
  We present derivations of the contagion condition for a range of spreading mechanisms
on families of generalized random networks and bipartite random networks.
We show how the contagion condition can be broken into three
elements, two structural in nature, and the third a meshing of
the contagion process and the network.
The contagion conditions we obtain reflect
the spreading dynamics in a clear, interpretable way.
For threshold contagion, we discuss results for all-to-all and random network versions
of the model, and draw connections between them.

\end{abstract}


\maketitle

\section{Introduction}
\label{sec:basiccontagion.introduction}

Given a local contagion mechanism acting on a random network,
and a seed set of nodes $\seedset$, we would like
to know the answers to a series of increasingly
specific questions:
\begin{enumerate}
\item[Q1:]
  Is a global spreading event possible?
  We'll define a ``global spreading event'' as one
  that reaches a non-zero fraction of a network
  in the infinite limit.
\item[Q2:]
  If a global spreading event is possible, what's the probability of one occurring?
\item[Q3:]
  What's the distribution of final sizes for all spreading events?
\item[Q4:]
  Global or not, how does the spreading from the seed set $\seedset$ unfold in time?
\end{enumerate}

Now, if we know the full time course of a spreading event (Q4)
(see~\cite{gleeson2007a}),
we evidently will be able to answer questions 1, 2, and 3.
We might be tempted to take on only the more challenging analytical work
and call it day (or appropriate time frame of suffering required).
But it turns out to be useful to address each question separately

While we will take on these questions for simple model distillations only,
their real-world counterparts
are some of the most important ones we face.
What's the probability that a certain fraction of
a population will contract influenza?
Could an ecosystem collapse?
Indeed, the biggest question for many systems is:

\begin{enumerate}
\item[Q5:]
  If we have limited knowledge of a network and limited control, how do we optimally
  facilitate or prevent spreading~\cite{lazarsfeld1954a,watts2007a}?
\end{enumerate}

In this chapter, we'll focus on Q1, determining the \textit{contagion condition}
for a range of contagion processes on random networks including bipartite ones.
We will do so by plainly encoding the course of the spreading process itself
into the contagion condition.

We will take the basic contagion mechanism to be one for which
there are node states: Susceptible (S) and Infected (I).
We will also prevent nodes from recovering or becoming susceptible;
once nodes are infected, they remain so.
In mathematical epidemiology, such models
are referred to as SI, where S stands for Susceptible and I for Infected.
Two other commonly studied models are SIR and SIRS,
where a recovered immune state R is allowed for both
and the possibility of cycling in the latter.

For the most part, we will be considering infinite random networks.
If needed, we will define such networks as the limit of a
one parameter family of networks
(e.g., \erdosrenyi\ networks with increasing $N$ and
mean degree held constant).
As a rough guide for simulations, using around $N=10^4$ nodes
is typically sufficient for yield results that visually
conform well to theoretical ones (e.g., fractional size of
the largest component in \erdosrenyi\ networks).

\section{Elements of simple contagion on random networks}
\label{sec:basiccontagion.elements}

\begin{figure}[tbp!]
  \includegraphics[width=0.5\columnwidth]{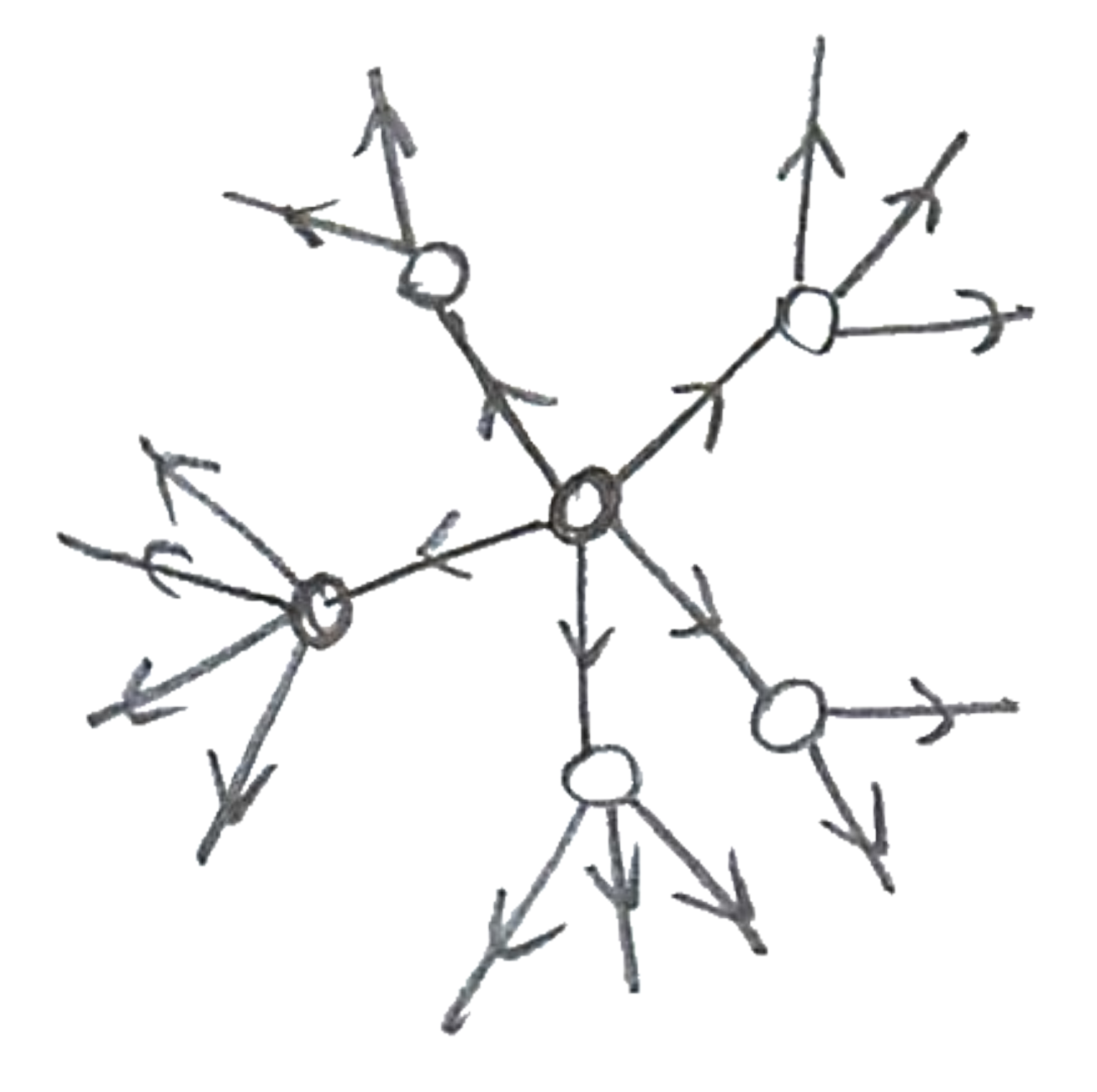}
  \caption{
    Random networks are locally pure branching structures.
    For the initial stages of the spread shown, nodes can only
    experience the infection from a single neighbor.
    For spreading to take off from a simple seed,
    the network must contain a connected macroscopic critical mass network $\Omega_{\crit}$
    of nodes susceptible to
    a single neighbor becoming infected.
  }
  \label{fig:basiccontagion.spreadingexample}
\end{figure}

The key feature of random networks for spreading is
that they are locally pure branching structures.
This remains true for a large number of variations
on random networks such as correlated random
networks and bipartite affiliation graphs.
Successful spreading away from a single seed (which could be one 
of many seeds) can only occur if nodes are susceptible when
just one of their neighbors is infected
(see Fig.~\ref{fig:basiccontagion.spreadingexample}).
We will refer to these easily susceptible nodes
as critical nodes
(called vulnerable nodes in~\cite{watts2002a}).
Denoting a network's entire node set as $\Omega$,
global spreading will only be possible
if there is a connected subnetwork of
critical nodes that forms a giant component,
the critical mass network $\Omega_{\crit}$.

This set of critical nodes behaves in the same
way as a critical mass one does for collective action~\cite{olson1971a,granovetter1978a,oliver1985a,oliver1993a}
but there is now an internal dynamic.
If one node is infected within
the critical mass network $\Omega_{\crit}$,
then spreading to some fraction of the critical mass network
and beyond is possible, depending on the probablistic
nature of the contagion process.

There are two other subnetworks that need
to be characterized to understand spreading on random networks.
First, containing the critical mass network and
all non-critical nodes connected to the critical mass network
is the triggering component, $\Omega_{\trig}$.
Knowledge of this structure is required
to determine the probability of a global spreading event~\cite{harris2014a}.
Second, we have $\Omega_{\textnormal{final}}$
which is the extent of infection realized for any spreading event.
For random networks,
the distribution of the fractional size of $\Omega_{\textnormal{final}}$ 
will be either unimodal (the contagion process always succeeds)
or bimodal (initial failure is possible).

\begin{figure}[tbp!]
  \includegraphics[width=\columnwidth]{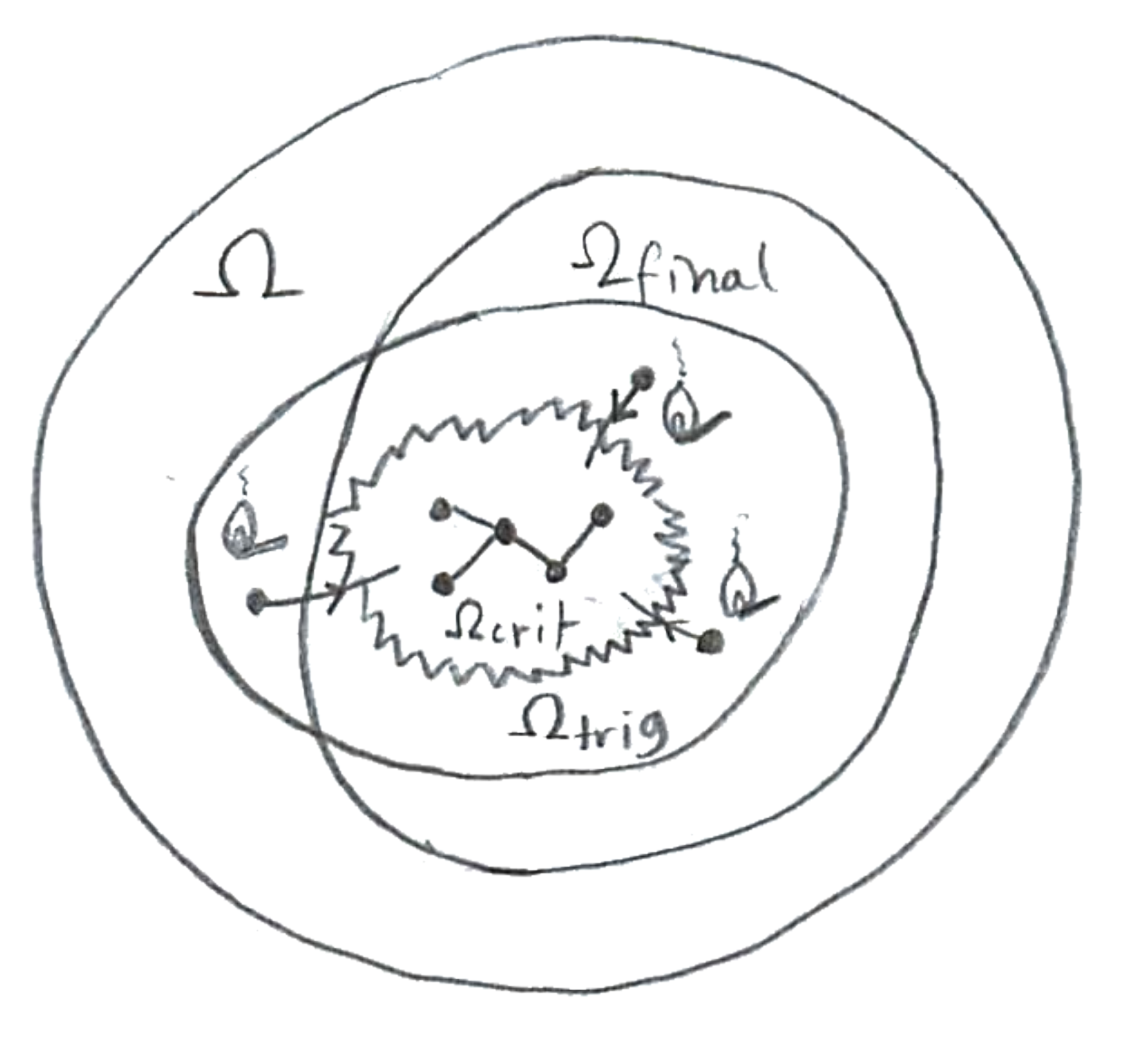}
  \caption{
    One possible arrangement of the three essential subnetworks
    for a contagion process on a random network:
    the critical mass network $\Omega_{\textnormal{crit}}$,
    the triggering component $\Omega_{\trig}$,
    and the final extent of a global spreading process,
    $\Omega_{\textnormal{final}}$.
    In general,
    $
    \Omega_{\textnormal{crit}} 
    \subset
    \Omega_{\trig}
    $,
    $
    \Omega_{\textnormal{crit}} 
    \subset
    \Omega_{\textnormal{final}}
    $,
    and
    $
    \Omega_{\trig},
    \Omega_{\textnormal{final}} 
    \subset
    \Omega.
    $
  }
  \label{fig:basiccontagion.bigspreadingpicture}
\end{figure}

If Fig.~\ref{fig:basiccontagion.bigspreadingpicture},
we show how the
three subnetworks
$\Omega_{\crit}$,
$\Omega_{\trig}$,
and
$\Omega_{\textnormal{final}}$
potentially overlap.
A global spreading event is only possible
if $\Omega_{\crit}$ takes up a non-zero
fraction of the network.
Some limiting cases allow for surprising kinds of robust-yet-fragile contagion,
such as 
$\Omega_{\crit}$
being vanishingly small
while any successful infection spreads to the full network~\cite{watts2002a}.

\section{The Contagion Condition}
\label{fig:basiccontation.contagioncondition}

We would like to devise some kind of general,
quick test algorithm
into which we would be able to feed any contagion mechanism and any network,
whether constructed or real.
Such an algorithm would generate what we'll call a Contagion Condition,
and would only be worthwhile if it avoided simulating all
possible spreading events and instead computed a composite test statistic.
Upon running a system through our algorithm
we would simply receive a ``Yes'' or ``No''.
Scaling up, we could then test an array of systems in parallel
and for the ``Yes'' responses, we would proceed to
explore those systems in detail
(e.g., those cities which are susceptible to Zombie outbreaks~\cite{munz2009a}).

\subsection{Contagion condition for one-shot spreading processes}

For random network models, our test algorithm can be
formulated in a physically-minded way.
We will step through the building of the contagion condition for one-shot, permanent infection spreading
on generalized, uncorrelated random networks and then expand from there.

By one-shot spreading, we mean that each newly infected node
has one chance in the next time step to infect its uninfected neighbors.
That is, if node $i$ fails to infect a specific neighbor $i'$,
then $i$ cannot attempt to infect $i'$ again in any following time step.
Permanent infection means that nodes do not recover.

For a node $i$ with degree $k$,
we will write $i$'s probability of infection given
$j$ of its neighbors are infected
as $\infprob_{kj}$.
While our focus on the initial spread on random networks means
we need only consider the probability nodes are infected
by one of their neighbors, $\infprob_{k1}$,
we must consider the response to multiple simultaneous infections
for later stages of global spreading on random networks~\cite{gleeson2007a,gleeson2008a},
more complicated contagion mechanisms,
and, more importantly if we care about the real world,
networks with non-zero clustering~\cite{watts1998a,newman2003a}.

As is often the case with networks, we open up better
ways to understand and explain phenomena if we focus
on edges rather than nodes.
This is not entirely natural as for many problems
we are ultimately concerned with how nodes behave and,
for contagion especially,
we can readily map ourselves directly onto individual nodes
(will my next movie fail?).
But once we lose this anchoring and shift to thinking first
about edges with nodes in the background, clearer paths emerge.

So, instead of framing spreading as rooted in node infection rates,
we consider the dynamics of infected edges.
For our purposes, an infected edge will be one emanating from an infected node,
and we will have to consider direction even for undirected networks.

We need to determine one number for our system, what we'll
call the gain ratio, $\gainratio$~\cite{dodds2011b}.
We define $\gainratio$ as
the expected number of newly infected edges that will be generated
by a single infected edge leading to an uninfected node.
(In epidemiology, the gain ratio would be equivalent to the reproduction number, $R_{0}$.)

For the moment, let's assume we have computed $\gainratio$
for a system.
Because sparse random networks are locally pure branching
structures
(see Fig.~\ref{fig:basiccontagion.spreadingexample}),
the spread emanating from a single seed will also
be a simple branching one.
Early on, there will be no interactions
between any two newly infected edges
leading to the same uninfected node.

The fraction of newly infected edges at time $t$,
$\edgeinffreqplain(t)$,
must then follow an elementary evolution:
\begin{equation}
  \edgeinffreqplain(t)
  =
  \gainratio
  \edgeinffreqplain(t-1).
  \label{eq:basiccontagion.spreadingratio}
\end{equation}
The subscript for the count
$\edgeinffreqbare$
will indicate the edge's type
which for our initial system is irrelevant,
hence $(\cdot)$.

The early growth will therefore be exponential with
\begin{equation}
  \edgeinffreqplain(t)
  =
  \gainratio^{t}
  \edgeinffreqplain(0),
  \label{eq:basiccontagion.spreadingsolution}
\end{equation}
where
$\edgeinffreqplain(0)$
equals the degree of
the seed node.
We might guess that we can write down
the exact evolution as 
$
\edgeinffreqplain(t)
=
\gainratio^{t}
\edgeinffreqplain(0),
$
but the initial step is sneakily different.
Well get to this issue later on.

Global spreading will evidently be possible only if
\begin{equation}
  \gainratio > 1,
  \label{eq:basiccontagion.spreadingcondition}
\end{equation}
and this very simple criterion will be our Contagion Condition.

The above equations maintain the same form if we
consider not one seed but
a random seed set taking up a non-zero fraction of the random network.
Writing
$\edgeinfprob_{t}$
as the fraction of edges emanating from newly infected
nodes at time $t$,
we have,
again for the initial phase of spreading:
\begin{equation}
  \edgeinfprob_{t}
  =
  \gainratio
  \edgeinfprob_{t-1},
  \label{eq:basiccontagion.spreadingratiofrac}
\end{equation}
which leads to
\begin{equation}
  \edgeinfprob_{t}
  =
  \gainratio^{t}
  \edgeinfprob_{0}.
  \label{eq:basiccontagion.spreadingsolutionfrac}
\end{equation}

We now determine the gain ratio $\gainratio$
for the simple class of one-shot contagion on
random network systems.
In doing so, we show that the Contagion Condition
is worthwhile beyond being a simple diagnostic
as,
with the right treatment,
it can be also seen to carry physical intuition.

In determining $\gainratio$,
there are three (3) pieces to consider:
two are structural and a function
of the network,
and the third couples the contagion
mechanism to the network.

\begin{enumerate}
\item 
  We start on an edge that has just become infected
  and look toward the uninfected node that has
  now become exposed.  
  The properly normalized probability that this node
  has degree $k$ is
  \begin{equation}
    Q_k
    =
    \frac{kP_k}{\tavg{k}}
    \label{eq:basiccontagion.Qk}
  \end{equation}
  because each degree $k$ node can be reached
  along its $k$ edges.
  This skewing of the degree distribution is
  a result of some renown as it
  drives the Simon-like rich-get-richer models of network
  growth of Price~\cite{price1965a,price1976a}
  and Barab\'{a}si and Albert~\cite{barabasi1999a},
  and also
  underlies the friendship paradox
  and its generalizations~\cite{eom2014a,momeni2016a}:
  Your friends are quite likely to be different from you,
  and often in disappointing ways such as
  by having more friends or wealth on average.
\item
  Second, we have the action of contagion mechanism.
  As have already defined,
  with probability $\infprob_{k1}$ 
  the node of degree $k$ is infected by the single incoming infected edge.
  With probability $1-\infprob_{k1}$,
  the infection fails.
\item
  Depending on whether or not the infection
  is successful,
  we know that 
  in the next time step the contagion
  mechanism will generate either 0 or $k-1$ new
  infected edges.
\end{enumerate}

Putting these pieces together, we have 
\begin{gather}
  \gainratio
=
\sum_{k=0}^{\infty}
\underbrace{
  \frac{kP_k}{\tavg{k}}
}_{
  \mbox{ \scriptsize
    \begin{tabular}{l}
      prob. of\\
      connecting to \\
      a degree $k$ node\\
    \end{tabular}
  }
}
\bullet
\underbrace{
  \infprob_{k1}
}_{
  \mbox{\scriptsize
    \begin{tabular}{l}
      Prob. of \\
      infection
    \end{tabular}
  }
}
\bullet
\underbrace{
  (k-1)
}_{
  \mbox{\scriptsize
    \begin{tabular}{l}
      \# outgoing \\
      infected \\
      edges
    \end{tabular}
  }
}
\nonumber \\
+ \
\sum_{k=0}^{\infty}
\underbrace{
  \frac{kP_k}{\tavg{k}}
  }_{
  \mbox{ \scriptsize
    \begin{tabular}{l}
      prob. of\\
      connecting to \\
      a degree $k$ node\\
    \end{tabular}
  }
}
\bullet
\underbrace{
  (1-\infprob_{k1})
}_{
  \mbox{\scriptsize
    \begin{tabular}{l}
      Prob. of \\
      no infection
    \end{tabular}
  }
}
\bullet
\underbrace{(\ 0\ )}_{
  \mbox{\scriptsize
    \begin{tabular}{l}
      \# outgoing \\
      infected \\
      edges
    \end{tabular}
  }
}
\label{eq:basiccontagion.gainratiocalc}
\end{gather}
The second piece evaporates and we have
our contagion condition:
\begin{equation}
  \alertb{
    \
    \gainratio
    =
    \sum_{k=0}^{\infty}
    \frac{kP_k}{\tavg{k}}
    \bullet
    \infprob_{k1}
    \bullet
    (k-1)
    > 1.
    \
}
\label{eq:basiccontagion.contagioncondition-oneshot}
\end{equation}
Again, the value here is that this structure of $\gainratio$
encodes the contagion mechanism in a clear way.
As such, we resist any urge to rearrange the form
of~\Req{eq:basiccontagion.contagioncondition-oneshot}
for a more elegant form.
As we move to more general systems,
the three part form of two pieces for the network and
one for the contagion mechanism will be maintained,
and the criterion
of a single number exceeding unity,
$\gainratio>1$,
will elevate
to being the largest
eigenvalue of a gain ratio matrix
exceeding unity.

We now move through a few examples of other kinds
of systems involving
contagion mechanisms acting on network structures.

\subsection{Contagion condition for multiple-shot spreading processes}

We have presumed a one-shot contagion process
in our derivation
of~\Req{eq:basiccontagion.contagioncondition-oneshot}.
In loosening this restriction to spreading
processes that may involve repeated attempts
to infect a node with the possible recovery of the
infected node allowed as well,
we can compute $\infprob_{k1}$
as the long term probability of
infection.
The form of gain ratio remains the same
and therefore so does the contagion condition
given in~\Req{eq:basiccontagion.contagioncondition-oneshot}.

\subsection{Remorseless spreading and the giant component condition}

We step back from contagion momentarily to show
that we can also determine
whether or not a random network has a giant component.
This is now a structural test absent any processes.
A network will have a giant component
if it is, on average, locally expanding.
That is, if we travel along a randomly chosen edge,
we will reach a node which has, on average, more than
one other edge emanating from it.
But this is just a remorseless version
of our one-shot contagion mechanism, one where
infection always succeeds, i.e., $\infprob_{k1}=1$.

Setting $\infprob_{k1}=1$
in~\Req{eq:basiccontagion.contagioncondition-oneshot},
we have the giant component condition:
\begin{equation}
\gainratio = 
\sum_{k=0}^{\infty}
\frac{kP_k}{\tavg{k}}
\bullet
(k-1)
>
1,
\label{eq:basiccontagion.giantcomponent}
\end{equation}
where we have again used the
physical sense of a gain ratio.

\subsection{Simple contagion on generalized random networks}

If $\infprob_{k1}=\infprob<1$,
A fraction (1-$\infprob$) of all edges will not transmit 
infection, 
and the contagion condition becomes
\begin{equation}
 \gainratio = 
\sum_{k=0}^{\infty}
\frac{kP_k}{\tavg{k}}
\bullet
\infprob
\bullet
(k-1)
> 1.
\label{eq:basiccontagion.bondperc}
\end{equation}
This is a bond percolation model~\cite{stauffer1992a},
and~\Req{eq:basiccontagion.bondperc}
can be seen as a giant component condition for
a network with (1-$\infprob$) of its edges removed.
The resultant network has a degree distribution
$
\tilde{P}_k
=
\infprob^k
\sum_{i=k}^{\infty}
\binom{i}{k}
(1-\infprob)^{i-k}
P_i,
$
and evidently, as $\infprob$ decreases, only increasingly
more connected networks will be able to facilitate spreading.

\subsection{Other routes to determining the contagion and giant component conditions}

There are many other ways to arrive at the contagion condition
in~\Req{eq:basiccontagion.contagioncondition-oneshot}
and the giant component condition
in~\Req{eq:basiccontagion.giantcomponent}.
The path taken affects the form of the condition
and may limit understandability~\cite{dodds2011b}.
For example, the giant component condition was
determined by Molloy and Reed~\cite{molloy1995a} in 1995
and presented as
\begin{equation}
  \sum_{\kplain=0}^{\infty}
  \kplain
  (\kplain - 2)
  P_{\kplain}
  >
  0.
  \label{eq:basicontagion.molloyreed}
\end{equation}
While equivalent to~\Req{eq:basiccontagion.giantcomponent},
the framing of local expansion is obscured.

For a simple spreading mechanism
with $\infprob_{k1}=\infprob$,
Newman~\cite{newman2003a}, for example, used
generatingfunctionology methods~\cite{wilf2006a}
to first determine
the average size of finite components and then
find when this quantity diverged.
For Granovetter's social contagion threshold model on
random networks~\cite{granovetter1978a},
Watts took the same approach~\cite{watts2002a}.
This size divergence is a hallmark of
phase transitions in statistical mechanical
systems in general, and while it can be
used to find the critical point,
doing so would ideally be at the level of a consistency check.

For the giant component condition,
a somewhat more direct approach using generating functions~\cite{newman2001b}
is based on the probability distribution
that the node at the randomly chosen
end of a randomly chosen edge has $k$ other edges
is
\begin{equation}
R_{k}
=
Q_{k+1}
=
\frac{1}{{\tavg{k}}}
(k+1)P_{k+1}.
\label{eq:basicontagion.Rk}
\end{equation}
Writing the generating function for the degree
distribution as
$
F_{P}(x)
=
\sum_{k=0}^{\infty}
P_k
x^k
$,
we have
$
F_{R}(x)
=
F'_{P}(x)/F'_{P}(1),
$
where we have used
$\tavg{k}=F'_{P}(1)$,
an elementary result for determining averages with generating functions~\cite{wilf2006a}.
The average number of other edges found at a randomly-arrived-at node
is
$
F'_{R}(1)
=
F''_{P}(1)/F'_{P}(x)
=
\frac{\tavg{k(k-1)}}
     {\tavg{k}}.
$
This is exactly our gain ratio and we now
have
\begin{equation}
  \frac{
    \tavg{k(k-1)}
  }{
    \tavg{k}
  }
  >
  1
  \label{eq:basiccontagion.gfgainratio}
\end{equation}
for the giant component condition.
Again, while Eqs.~\ref{eq:basiccontagion.giantcomponent}
and~\ref{eq:basiccontagion.gfgainratio} are equivalent,
the latter does not have an immediate physical
interpretation---it's just a condition.

\subsection{Simple contagion on generalized directed random networks}
\label{subsec:generalcontagion.generalizedrandom}

\begin{figure}[tbp!]
  \includegraphics[width=0.5\columnwidth]{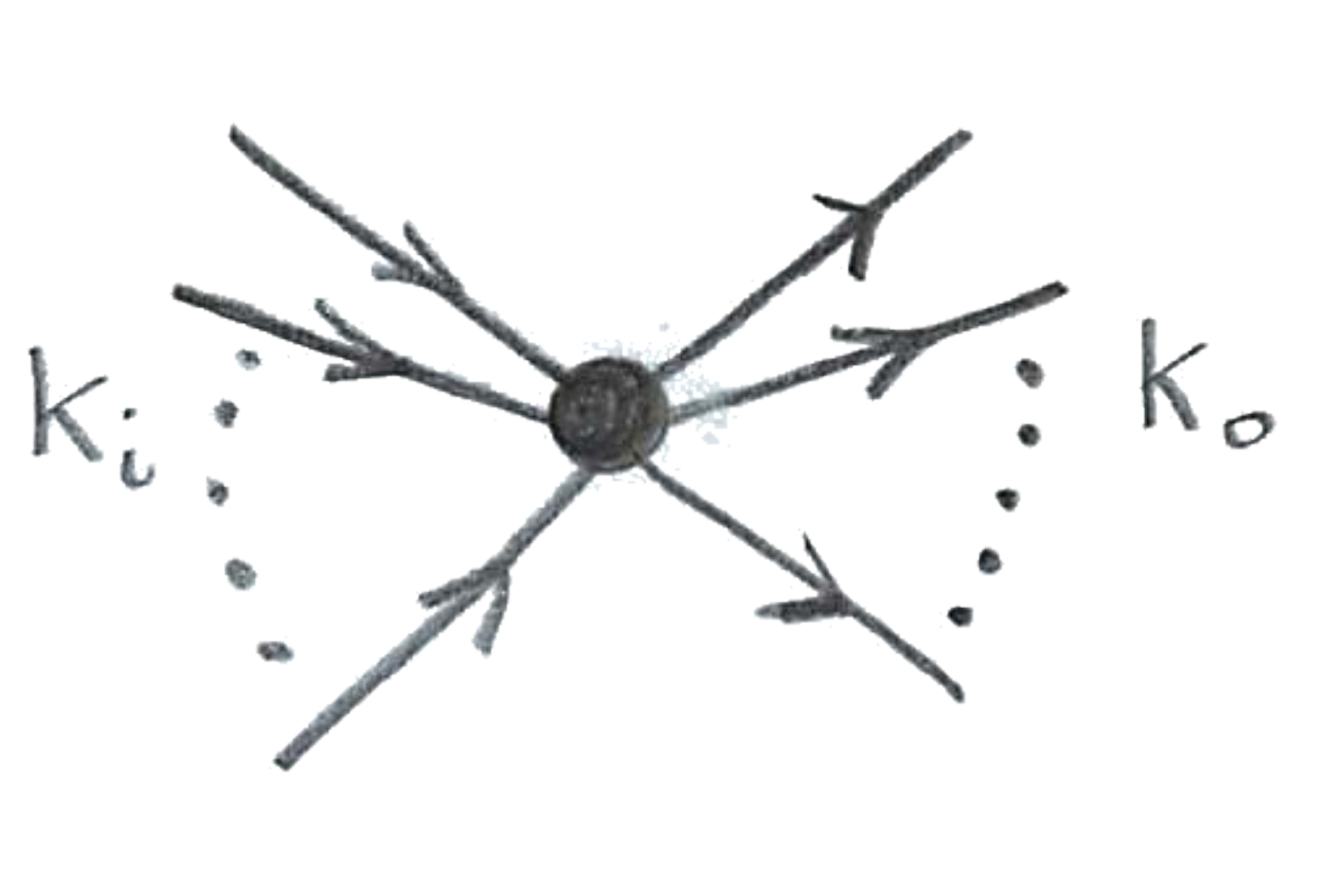}
  \caption{
    For general directed networks,
    a node has
    $\kin$ incident edges
    and $\kout$ emanating edges
    governed by a
    a joint distribution $P_{\kin,\kout}$.
  }
  \label{fig:basiccontagion.directed}
\end{figure}

For purely directed networks,
we allow each node to have 
an in-degree $\kin$ and an out-degree $\kout$
with probability
$P_{\kin,\kout}$ (see Fig.~\ref{fig:basiccontagion.directed}).
The same arguments that gave us 
\Req{eq:basiccontagion.contagioncondition-oneshot}
now end with:
\begin{equation}
 \gainratio = 
\sum_{\kin=0}^{\infty}
\sum_{\kout=0}^{\infty}
\frac{\kin P_{\kin,\kout}}{\tavg{\kin}}
\bullet
\infprob_{\kin,1}
\bullet
\kout
> 1.
\label{eq:basiccontagion.bondpercdirected}
\end{equation}
The three components of the contagion condition
have the same interpretation as before.

\subsection{Simple contagion on mixed, correlated random networks}
\label{subsec:generalcontagion.mixed}

\begin{figure}[tbp!]
  \includegraphics[width=0.5\columnwidth]{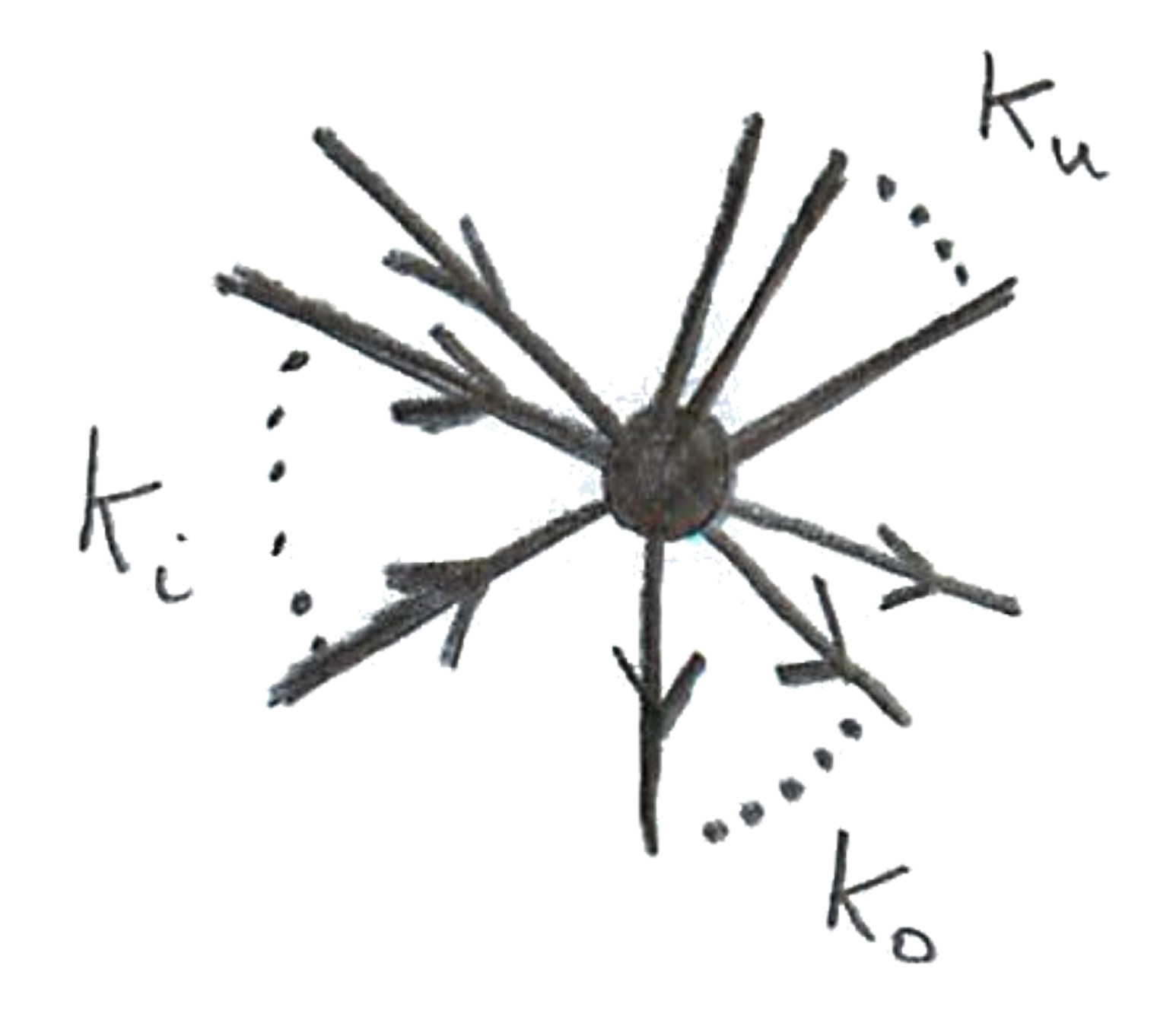}
  \caption{
    Nodes in mixed random networks
    have $\kbid$ undirected edges,
    $\kin$ incident edges,
    and $\kout$ emanating edges.
    Node degree is represented by
    the vector
    $
    \veck = [\ \kbid \  \kin \  \kout\ ]^{\textnormal{T}}
    $
    and
    degrees are sampled from
    a joint distribution
    $P_{\veck}$.
  }
  \label{fig:basiccontagion.mixednetork}
\end{figure}

We jump to a more complex possibility of
mixed random networks with a combination of
directed and undirected (or bidirectional) edges
as well as arbitrary degree-degree correlations between
nodes, as introduced in ~\cite{boguna2005a}.
      
Nodes may have three types of edges:
$\kbid$ undirected edges,
$\kin$ incoming directed edges,
and
$\kout$ outgoing directed edges.
The degree distribution is now a function
of a three-vector:
\begin{equation}
    P_{\veck}
    \ \mbox{where} \
    \veck = [\ \kbid \ \  \kin \ \ \kout\ ]^{\textnormal{T}}.
  \label{eq:basiccontagion.vectordegree}
\end{equation}
As for directed networks, we require in- and out-degree
averages to match up:
$
\tavg{\kin}
=
\tavg{\kout}.
$
We add two point correlations per~\cite{boguna2005a,dodds2011b}
through three conditional probabilities:
\begin{itemize}
\item 
  $
  \Probbid(\veck\,|\, \veck')
  $
  = 
  probability that an undirected edge leaving 
  a degree $\veck'$ nodes arrives at a degree $\veck$ node.
\item
  $
  \Probin(\veck\,|\, \veck')
  $
  = probability that an edge leaving 
  a degree $\veck'$ nodes arrives at a degree $\veck$ node
  is an in-directed edge
  relative to the destination node.
\item
  $
  \Probout(\veck\,|\, \veck')
  $
  = probability that an edge leaving 
  a degree $\veck'$ nodes arrives at a degree $\veck$ node
  is an out-directed edge
  relative to the destination node.
\end{itemize}

\begin{figure}
      \includegraphics[width=\columnwidth]{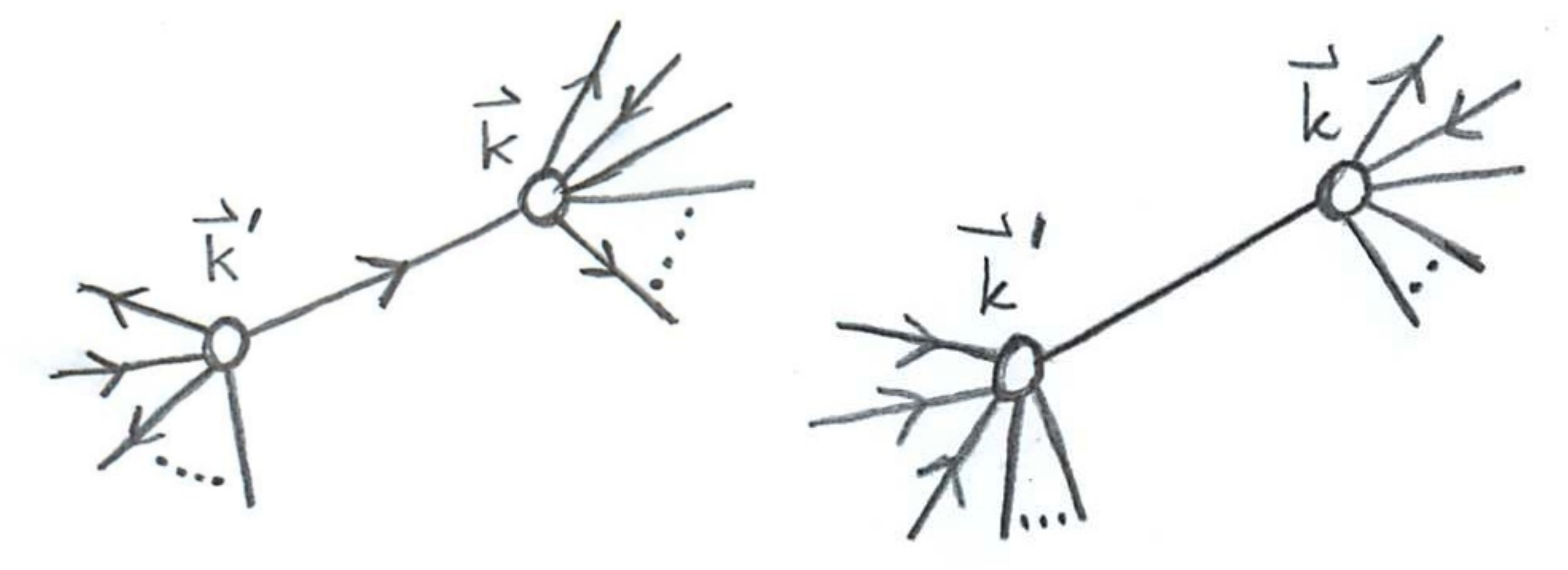}
  \caption{
    For mixed random networks,
    node degree correlations
    may be measured along
    undirected and/or directed edges.
  }
  \label{fig:basiccontagion.correlations}
\end{figure}

We now require more refined (detailed) balance
along both undirected and directed edges
(see Fig.~\ref{fig:basiccontagion.correlations}).
Specifically, we must have~\cite{boguna2005a,dodds2011b}:
$
\Probbid(\veck\, |\, \veck') 
\frac{\kbid' P(\veck')}{\tavg{\kbid'}}
=
\Probbid(\veck'\, |\, \veck) 
\frac{\kbid P(\veck)}{\tavg{\kbid}},
$
and
$
\Probin(\veck\, |\, \veck') 
\frac{\kout' P(\veck')}{\tavg{\kout'}}
=
\Probout(\veck'\, |\, \veck) 
\frac{\kin P(\veck)}{\tavg{\kin}}.
$
    
For all example systems so far, the gain ratio has been a single number.
For mixed random networks,
infections along directed edges may cause
infections along undirected edges and so on.
We will need to count undirected and directed
edge infections separately, the growth
of infections for a one-shot contagion process
will obey the following dynamic:
\begin{equation}
  \left[
    \begin{array}{c}
      f_{\veck}^{\textnormal{(\bidmark)}}(t+1) \\
      f_{\veck}^{\textnormal{(\outmark)}}(t+1)
    \end{array}
    \right]
  =
  \sum_{\veck'}
  \textbf{R}_{\veck\veck'}
  \left[
    \begin{array}{c}
      f_{\veck'}^{\textnormal{(\bidmark)}}(t) \\
      f_{\veck'}^{\textnormal{(\outmark)}}(t)
    \end{array}
    \right],
    \label{eq:basiccontagion.Rmatrixequation}
\end{equation}
where
we now identify a gain ratio tensor: 
\begin{eqnarray}
    \lefteqn{\textbf{R}_{\veck\veck'}
      = }
    \\
    & &
    \left[
    \begin{array}{ll}
      \Probbid(\veck\,|\, \veck')
      \bullet
      \infprob_{\veck\veck'}
      \bullet
      (\kbid-1)
      &
      \Probin(\veck\,|\, \veck')
      \bullet
      \infprob_{\veck\veck'}
      \bullet
      \kbid
      \\
      \Probbid(\veck\,|\, \veck')
      \bullet
      \infprob_{\veck\veck'}
      \bullet
      \kout
      &
      \Probin(\veck\,|\, \veck')
      \bullet
      \infprob_{\veck\veck'}
      \bullet
      \kout
    \end{array}
    \right].
    \nonumber
    \label{eq:basiccontagion.Rmatrix}
\end{eqnarray}
For a gain ratio matrix or tensor,
our contagion condition is now a test
of whether or not the largest eigenvalue exceeds 1.

\subsection{Contagion on correlated random networks with arbitrary node and edge types}
\label{subsec:basiccontagion.arbitrary}

\begin{figure}
  \includegraphics[width=0.75\columnwidth]{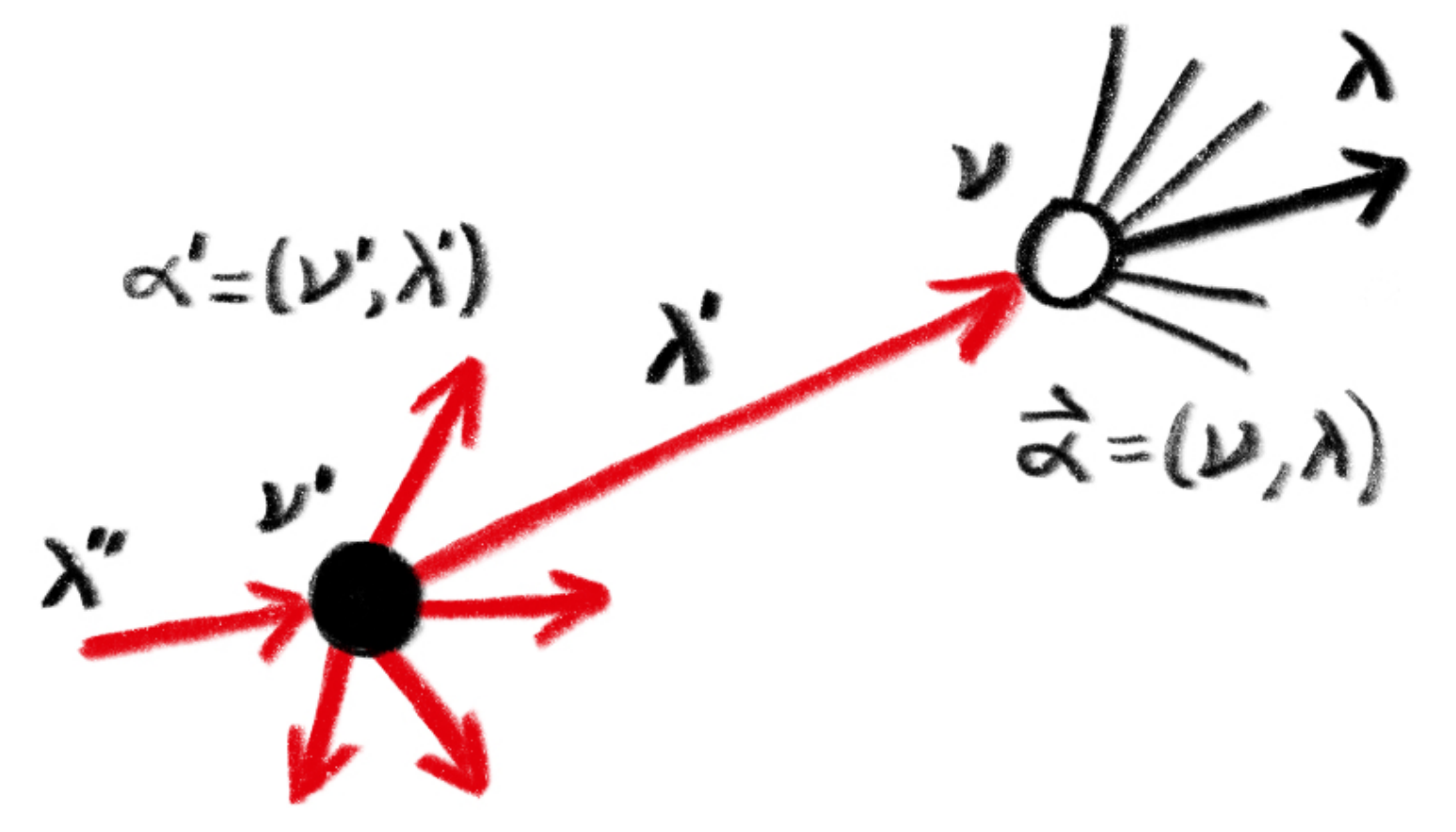}
  \caption{
    Element of a general correlated random network where edges and nodes
    may take on arbitrary characteristics.
    Node and edge type are specified
    as
    $\alpha=(\nu,\lambda)$.
  }
  \label{fig:basiccontagion.arbitrary}
\end{figure}

We make one last step of generalization for correlated random networks~\cite{dodds2011b}.
As per Fig.~\ref{fig:basiccontagion.arbitrary},
we allow arbitrary types of nodes and edges
along with arbitrary correlations between node-edge pairs.
For multi-shot contagion, we have
\begin{equation}
  f_{\vec{\alpha}}(d+1) 
  = 
  \sum_{\vec{\alpha}'}
  R_{\vec{\alpha} \vec{\alpha}'}
  f_{\vec{\alpha}'}(d)
  \label{eq:basiccontagion.Rmostgeneralequation}
\end{equation}
where
$R_{\vec{\alpha} \vec{\alpha}'}$
is the gain ratio matrix
and has the form:
\begin{equation}
R_{\vec{\alpha} \vec{\alpha}'}
=
P_{\vec{\alpha} \vec{\alpha}'}
\bullet
k_{\vec{\alpha} \vec{\alpha}'}
\bullet
\infprob_{\vec{\alpha} \vec{\alpha}'}.
  \label{eq:basiccontagion.Rmostgeneralform}
\end{equation}
Here,
\begin{itemize}
\item 
$P_{\vec{\alpha} \vec{\alpha}'}$
= conditional probability that
a type $\lambda'$ edge emanating from a type $\nu'$ 
node leads to a type $\nu$ node.
\item 
$k_{\vec{\alpha} \vec{\alpha}'}$
= potential  number of newly 
infected edges of type $\lambda$ 
emanating from nodes of type $\nu$.
\item 
$\infprob_{\vec{\alpha} \vec{\alpha}'}$
= probability that
a type $\nu$ node is eventually infected by
a single infected type $\lambda'$ link arriving from a 
neighboring node of type $\nu'$.
\end{itemize}

Finally, we can write down our generalized contagion condition as:
\begin{equation}
\max{|\mu|} : 
\mu 
\in 
\sigma
\left(
\gainratio
\right)  
> 1,
\label{eq:basiccontagion.Rmostgeneral-contagion-condition}
\end{equation}
where
$\sigma(\gainratio)$
denotes
the eigenvalue spectrum of $\gainratio$.

\subsection{Simple contagion on bipartite random networks}
\label{subsec:generalcontagion.bipartite}

\begin{figure*}[tbp!]
    \centering
  \includegraphics[width=\textwidth]{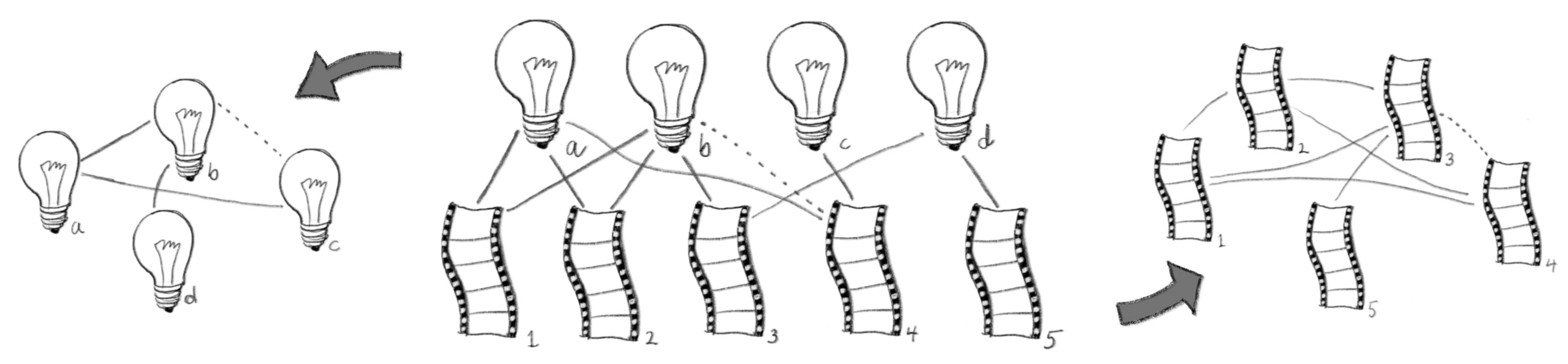}
  \caption{
    Example of a bipartite affiliation network and the induced networks.
    Center: A small story-trope bipartite graph.
    The induced trope network and the induced story network
    are on the left and right.
    The dashed edge in the bipartite affiliation network
    indicates
    an edge added to the system,
    resulting in the dashed edges being added to
    the two induced networks.
  }
  \label{fig:basiccontagion.bipartitelayout}
\end{figure*}

\begin{figure}[tbp!]
    \centering
  \includegraphics[width=\columnwidth]{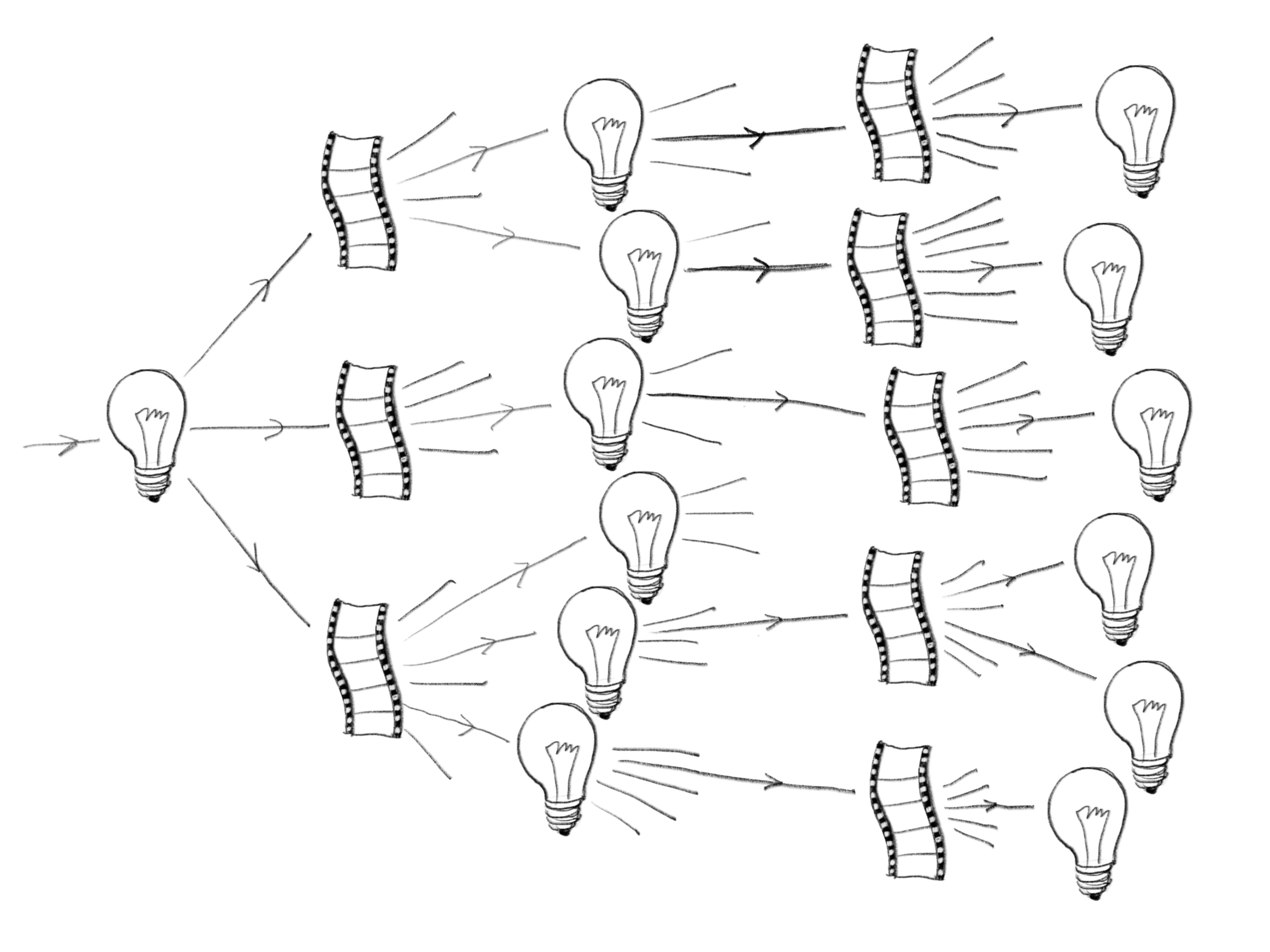}
  \caption{
    Spreading on a random bipartite network can be seen as bouncing
    back and forth between the two connected populations.
    The gain ratio for simple contagion on a bipartite random network
    is the product of two gain ratios as shown
    in~\Req{eq:basiccontagion.bipR}.
  }
  \label{fig:basiccontagion.bipartitespread}
\end{figure}

Bipartite networks (or affiliation graphs) connect two
populations through some association, and induce networks
within each population~\cite{newman2001b,ahn2011a,teng2012a,hidalgo2007a,goh2007a,garcia-perez2014a}.
Bipartite structures and variants are natural
representations of many real networked systems
with a classic example being boards and directors.
The induced distributions are formed by connecting all
pairs of boards that share at least one director and
all pairs of directors that belong to the same board.

Base models for real bipartite systems 
are random bipartite networks which are formed by randomly connecting
two populations with specified degree distributions.
Random bipartite networks are able to reproduce induced
degree distributions, which may be non-trivial in form~\cite{newman2001b}.

To help with our analysis, we'll consider
a random bipartite network between
stories and tropes~\cite{tvtropes2017a}.
Each story contains one or more trope,
and each trope is part of one more stories.
Stories sharing tropes are then linked
as are tropes found in the same story.
In Fig.~\ref{fig:basiccontagion.bipartitelayout},
we show a small example (center) along with
the induced trope-trope and story-story networks.

For spreading between
stories we may wish to imagine we're in
the BookWorld of the Thursday Next series~\cite{fforde2001a}.

We'll use this notation for our two inter-affiliated types:
$\rbone$ for stories
and 
$\rbtwo$ for tropes.

Consider a story-trope system with
$\rboneng$
denoting the number of stories,
$\rbtwong$ the number of tropes,
and
$m_{\rbone,\rbtwo}$
the number of edges connecting
stories and tropes.

Let's have some underlying distributions for numbers of
affiliations:
$P^{(\rbone)}_{k}$ (a story has $k$ tropes)
and
$P^{(\rbtwo)}_{k}$ (a trope is in $k$ stories).

Some bookkeeping arises with balance
requirements.
Writing 
$\tavg{k}_{\rbone}$
as the
average number of tropes per story,
and
$\tavg{k}_{\rbtwo}$
as the average number of stories containing a given trope,
we must have:
$
N_{\rbone} \cdot \tavg{k}_{\rbone}
=
m_{\rbone,\rbtwo}
=
N_{\rbtwo} \cdot \tavg{k}_{\rbtwo}.
$

Let's first get to the giant component condition
before talking about contagion.

Just as for random networks, we focus on edges begetting
edges, and we will need the distributions analogous
to $Q_{k}$, \Req{eq:basiccontagion.Qk}.
We randomly select an edge connecting 
a story $\rbone$ to a trope $\rbtwo$.
Traveling from the trope to the story,
we have that the 
probability the story $\rbone$ contains $k$
total tropes
is:
\begin{equation}
Q^{(\rbone)}_{k}
=
\frac{
  k P^{(\rbone)}_{k}
}{
  \sum_{j=0}^{N_\rbone} j P^{(\rbone)}_{j}
}
=
\frac{
  k P^{(\rbone)}_{k}
}{
  \tavg{k}_\rbone
}.
\label{eq:basicontagion.Rkone}
\end{equation}
Heading instead towards the trope $\rbtwo$, we find
the probability that the trope $\rbtwo$ is in $k$ total stories
is 
\begin{equation}
Q^{(\rbtwo)}_{k}
=
\frac{
  k P^{(\rbtwo)}_{k}
}{
  \sum_{j=0}^{N_\rbtwo} j P^{(\rbtwo)}_{j}
}
=
\frac{
  k P^{(\rbtwo)}_{k}
}{
  \tavg{k}_\rbtwo
}.
\label{eq:basicontagion.Rktwo}
\end{equation}

To determine the giant component condition
for the induced network of stories (to choose a side),
let's start with a randomly chosen edge
and travel from the story to the trope.
As shown starting on the left of Fig.~\ref{fig:basiccontagion.bipartitespread},
we hit the trope and then travel to
the other stories containing that trope.
This bouncing back and forth between
tropes and stories continues and because
the connections are random and if the
system is large enough, no story or
trope is returned to early on.
Just as for random networks, there
are no short loops (technically, finitely
many in the infinite limit).

We are thus able to depict the expanding
branching in Fig.~\ref{fig:basiccontagion.bipartitespread}
and we can see that the giant component condition
will involve the product of the gain ratio
for each distribution.
\begin{eqnarray}
  \lefteqn{
    \gainratio
    =
    \gainratio_{\rbone}
    \cdot
    \gainratio_{\rbtwo}
    =
  }
   \\
  &  &
  \left[
  \sum_{k=0}^{\infty}
  \frac{k P^{(\rbone)}_{k}}{\tavg{k}_{\rbone}}
  \bullet
  (k-1)
  \right]
  \left[
  \sum_{k=0}^{\infty}
  \frac{k P^{(\rbtwo)}_{k}}{\tavg{k}_{\rbtwo}}
  \bullet
  (k-1)
  \right]
  >
  1
  \nonumber
  \label{eq:basiccontagion.bipR}  
\end{eqnarray}
As for gain ratios for random networks
we can arrive at this result through
the use of generating
functions and other approaches.
Regardless of the path,
more mathematically pleasing variants
are always available such as~\cite{newman2001b}:
\begin{equation}
          \sum_{k=0}^{\infty}
          \sum_{k'=0}^{\infty}
          kk'(kk'-k-k') P^{(\rbone)}_{k} P^{(\rbtwo)}_{k'} = 0,
\label{eq:basiccontagion.bipRmath}  
\end{equation}
but, again, we have stripped the physics away.

Introducing a simple contagion can be
done as before by allowing tropes to
infect other tropes in the same story
(with probability
$\infprob_{k1}^{(\rbtwo)}$)
and stories to affect other stories if
they share a trope
(with probability
$\infprob_{k1}^{(\rbone)}$)
We adjust \Req{eq:basiccontagion.bipR}
to obtain:
\begin{eqnarray}
  \lefteqn{
    \gainratio
    =
    \gainratio_{\rbone}
    \cdot
    \gainratio_{\rbtwo}
    =
  }
   \\
  &  &
  \left[
  \sum_{k=0}^{\infty}
  \frac{k P^{(\rbone)}_{k}}{\tavg{k}_{\rbone}}
  \bullet
  \infprob_{k1}^{(\rbone)}
  \bullet
  (k-1)
  \right]
  \nonumber \\
  & &
  \times
  \left[
  \sum_{k=0}^{\infty}
  \frac{k P^{(\rbtwo)}_{k}}{\tavg{k}_{\rbtwo}}
  \bullet
  \infprob_{k1}^{(\rbtwo)}
  \bullet
  (k-1)
  \right]
  >
  1
  \nonumber
  \label{eq:basiccontagion.bipRmod}
\end{eqnarray}

\subsection{Threshold contagion on generalized random networks}

We turn to our last example: threshold contagion, an important simple
model of social contagion~\cite{schelling1971a,schelling1973a,schelling1978a,granovetter1978a,granovetter1983a,granovetter1986a,granovetter1988a,watts2002a}.
In basic threshold contagion models, all individuals observe the infection status
of their neighbors at each time step, and become infected if their
internal threshold is exceeded.
In the present and following section,
we will explore the contagion condition for threshold models
on all-to-all networks and random networks, and
examine the early course of a global spreading event
reflecting on the nature of early adopters.

In Granovetter's mean-field or all-to-all network version~\cite{granovetter1978a}, individuals are
always aware of the overall fraction of the population that is infected.
We write the fraction of the population that is infected at time
$t$ as $a_{t}$.
If we have a general threshold distribution \alert{$f(\phi)$},
then the fraction of the population whose threshold
will be exceeded at time $t$ and hence be infected
at time $t+1$ is:
\begin{equation}
  \phi_{t+1}
  =
  \int_{0}^{\phi_t} f(u) \dee{u}
  =
  \left. F(u) \right|_{0}^{\phi_t}
  =
  F(\phi_t)
  -
  F(0) 
  \label{eq:basiccontagion.granovetter}
\end{equation}
where  $F$ is the cumulative distribution of $f$
(if $F(0) > 0$, then the system has nodes that will
always be on regardless of the state of others).
Thus, we have system whose dynamics
are described by a map of the unit interval.
We are interested in small seeds for the mean-field version,
i.e.,
$\phi_{0} \rightarrow 0$.
In this limit, global spreading occurs if
(1) $F(0)>0$ meaning the population will always activate spontaneously, 
or 
(2) $\phi=0$ is a fixed point but is unstable (meaning $F(0)=0$ and $F'(0)>1$).
If $\phi=0$ is a stable fixed point (meaning $F(0)=0$ and $F'(0)<1$), then
spreading may still occur but not for vanishingly small seeds.
Perhaps surprisingly, the same process on a network may give
rise to spreading from a single seed, as we explain this in
the next section.

For the random network version due to Watts~\cite{watts2002a},
and again taking a general threshold distribution \alert{$f(\phi)$}
a degree $k$ node will be part of the critical mass network with probability:
\begin{equation}
\infprob_{k1} = \int_{0}^{1/k} f(\phi) \dee{\phi}.
\label{eq:basiccontagion.Bk1threshold}
\end{equation}
The gain ratio remains the same as the one
given in~\Req{eq:basiccontagion.contagioncondition-oneshot}.

We now link the contagion conditions for the all-to-all network
and random network versions of social contagion.

\subsection{Connecting the contagion condition for all-to-all and random networks for threshold contagion}
\label{subsec:basiccontagion.bounded_unbounded}

\begin{figure}[tbp!]
  \centering
  \includegraphics[width=.48\textwidth]{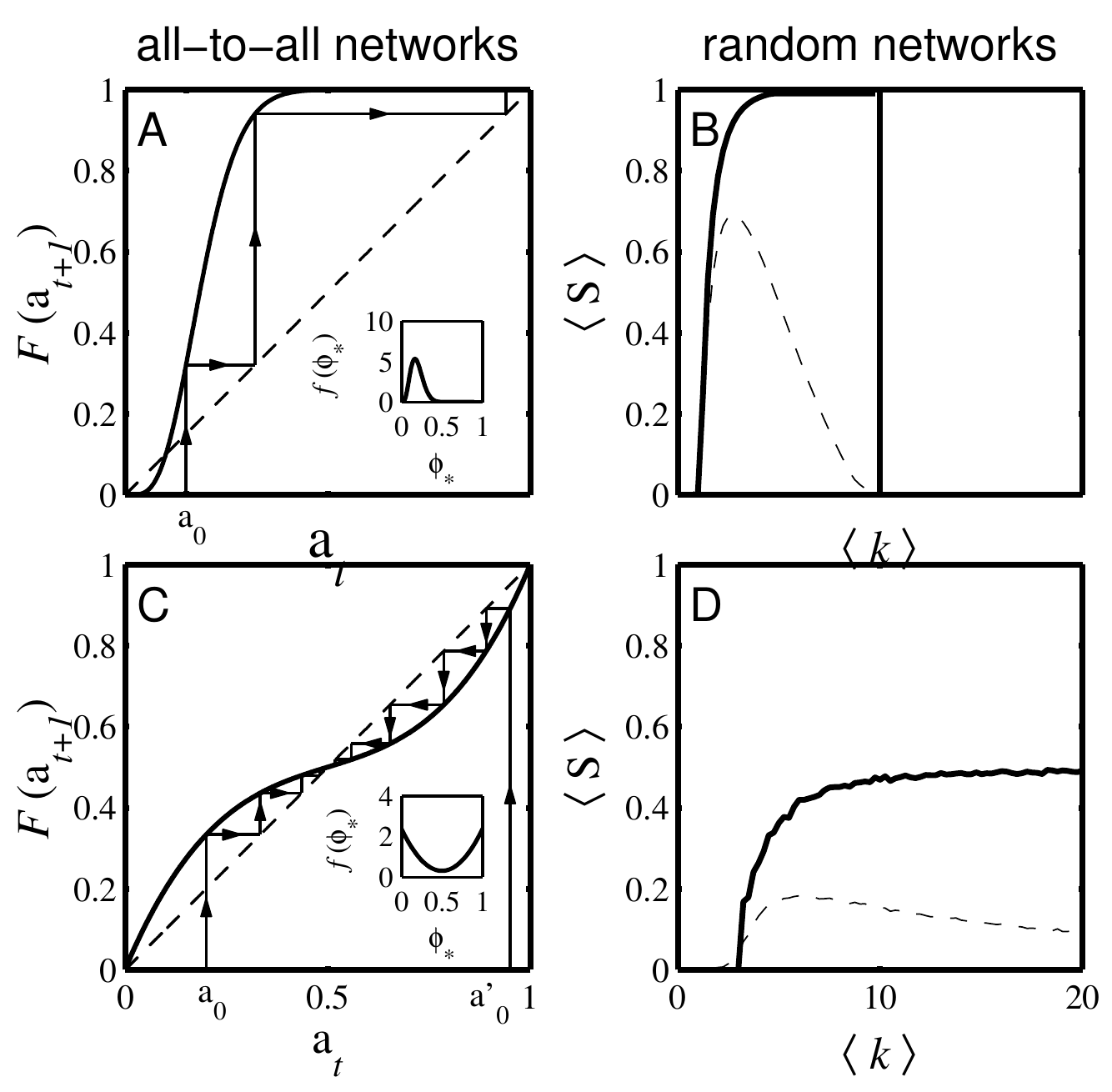}
  \caption{
    Plots comparing the behavior of the model on all-to-all networks (plots A and C)
    and random networks (B and D)
    for two different example threshold distributions.
    The insets to plots A and C show the two
    underlying threshold distributions, which are unimodal and bimodal
    respectively, and the corresponding cumulative distributions
    are presented in the main plots of A and C.
    Plots B and D show global spreading event intervals
    for random networks with the same threshold distributions
    as A and C respectively.
    The black lines in B and D indicate the average
    size of global spreading events that exceed $0.05N$, and the dashed lines the average
    size of the largest critical mass network (sizes are normalized by $N$)
    The threshold distribution in plot
    A leads to a bounded global spreading event interval on random networks while the distribution in plot C
    leads to an unbounded one.  In plot D, the average size of the largest critical mass network
    decays to 0 as $\tavg{k} \rightarrow \infty$.
    The results in plots B and D are derived from $10^3$ networks with $N=10^4$
    and one seed per network.
  }
  \label{fig:basiccontagion.cascwind}
\end{figure}

We make the simple observation that if we examine the threshold model's behavior
on a random network and allow the average degree $\tavg{k}$ to increase, then
the results will tend towards what we would observe on an all-to-all network.  
Since the limiting behavior of the contagion model on all-to-all networks is governed by the
presence or absence of fixed points of the cumulative threshold distribution
$F$, we are therefore able to state what the model's behavior on random networks
must tend towards as $\tavg{k}$ increases based solely on the form of $F$.

We consider two examples of threshold distribution $f$ to facilitate our
discussion.
First, for a general threshold distribution $f$, it is useful for us to
define a \textit{global spreading event interval}
as the range of $\tavg{k}$ for  which global spreading events are possible on a random network.
A simple example involving a bounded global spreading event interval and a non-trivial threshold
distribution $f$ is represented in Figs.~\ref{fig:basiccontagion.cascwind}A--B.
The main plot of Fig.~\ref{fig:basiccontagion.cascwind}A shows the 
cumulative distribution $F$, and the inset shows the threshold distribution $f$.
The all-to-all network model, \ref{fig:basiccontagion.cascwind}A,
exhibits a simple kind of critical mass behavior: the infection
level approaches unity if the initial activated fraction $\phi_0$
is above the sole unstable fixed point, or else it dies away.  Thus for
all-to-all networks, 
a small initial infection level will always fail to yield
global infection.  
For global spreading events to occur on all-to-all networks,
some alternative seeding mechanism (an advertising campaign, perhaps) 
must precede the word-of-mouth dynamics so as to create a sufficiently large $\phi_0$.  

By contrast, global spreading events can 
arise from a \textit{single} infected individual
in a sparse random network with exactly the same distribution of thresholds,
as shown in Fig.~\ref{fig:basiccontagion.cascwind}B.
The reason is that when individuals are connected to a limited number
of alters within a population,
the fraction of their neighbors that
are infected may now be nonzero and thus may exceed their threshold 
(in infinite all-to-all networks, this fraction is always 0 for finite seeds).
By effectively reducing the knowledge individuals have of the overall 
population---by increasing their ignorance---global spreading events become possible.
Related observations invoke pluralistic ignorance~\cite{kuran1991a,kuran1997a}
and the importance of small groups in facilitating collective action~\cite{olson1971a} 
by circumventing the free rider problem.

Thus, when the threshold distribution $f$ is fixed, we observe
a connection between the results for spreading
on all-to-all networks and random networks.
Bounded global spreading event intervals can only occur when the mean-field version exhibits
a critical mass property, i.e., when there exists a stable fixed point
at the origin $\phi=0$ (i.e., $F(0)=0$ and  $F'(0) < 1$).
We know this because no small seed will ever be able
to generate a global spreading event in the all-to-all case and that 
as the average degree of a random network increases, so too must
its similarity in behavior to that of all-to-all networks.
Furthermore, if there is a stable fixed point at the origin,
whether or not global spreading events are possible at all in any random network
depends on the global spreading event condition being satisfied.  In other words,
ignorance does not always help the spread of influence---some threshold distributions
never lead to the contagion condition being satisfied for any value of $\tavg{k}$.

\textit{Unbounded global spreading event intervals} arise when there are sufficient individuals who
will be vulnerable even if their degree is very high, i.e., 
when the threshold distribution has enough weight at or near $\phi=0$.
An example of an unbounded global spreading event interval is given in Fig.~\ref{fig:basiccontagion.cascwind}D
with the underlying threshold distribution and its cumulative shown
in Figs.~\ref{fig:basiccontagion.cascwind}C.
Since small seeds always take off in the all-to-all network version,
as network connectivity is increased, global spreading events continue to occur
and the global spreading event interval is unbounded.
The size of the largest critical mass network is nonzero 
for all finite $\tavg{k}$, though it tends to 0 in the limit $\tavg{k}\rightarrow \infty$.
For highly connected random networks, the final size of the global spreading event again depends on
the fixed points of $F$.  
For example, in Fig.~\ref{fig:basiccontagion.cascwind}B, global spreading events typically
reach the full size of the giant component which corresponds to an upper stable
fixed point of $F$ at $\phi=1$.  In Fig.~\ref{fig:basiccontagion.cascwind}D,
we see global spreading events only reach half the size of the population,
corresponding to the stable fixed point of $F$ at $\phi=1/2$.

We thus see that in moving from all-to-all networks to random networks, the
behavior of the threshold model changes qualitatively
in the sense that there exist threshold distributions
for which global spreading events started by a small seed cannot occur on an all-to-all network,
yet may occur on sparse, random networks.

\bigskip

\section{Concluding remarks:}
\label{sec.basicontagion.concludingremarks}

For any parameterized system that may afford global spreading,
the contagion condition is a fundamental criterion to determine.
We have outlined the contagion condition
for a range of 
contagion mechanisms acting on generalized random networks,
showing that the condition can be derived so as
to bear a clear imprint of the mechanism at work.
A similar approach can be used to lay out
the triggering probability of a global spreading event
in a readable form~\cite{harris2014a}.

While generating function approaches provided many of the first breakthroughs
giving the possibility and probability of spreading~\cite{newman2001b,watts2002b}
and have yielded powerful access to many other results,
they have tended to obscure the forms of the simplest ones such as the contagion condition.
These techniques are also inherently indirect
as they work by avoiding the giant component and characterizing
only finite ones.
Later work focusing on fractional seeds was able to go directly
into the giant component and determine not just the final size
but full time dynamics of global spreading events~\cite{gleeson2007a,gleeson2008a},
and we recommend continued pursuit of this line of attack going forward.

\vfill

\clearpage

\end{document}